\documentclass[11pt]{article}
\usepackage[english]{babel}
\usepackage[a4paper,top=2cm,bottom=2cm,left=3cm,right=3cm,marginparwidth=1.75cm]{geometry}

\usepackage{amsmath, amssymb}
\usepackage{graphicx}
\usepackage{hyperref}
\hypersetup{colorlinks=true, allcolors=blue}

\usepackage{booktabs}
\usepackage{multirow}
\usepackage{array}
\usepackage{tabularx}
\usepackage{subcaption}
\usepackage{float}

\usepackage[authoryear,round]{natbib}

\title{Detecting Historical Turning Points in Italian Media: A Complex Systems Approach to a Diachronic News Corpus}
\author{
  Dario Zarcone\thanks{Department of Physics and Chemistry, University of Palermo, Palermo, Italy} \thanks{Corresponding author: dario.zarcone01@unipa.it} \and
  Salvatore Miccichè\footnotemark[1] \and
  David Sánchez\thanks{Institute for Cross-Disciplinary Physics and Complex Systems (IFISC), UIB-CSIC, Palma de Mallorca, Spain}
}

\begin{document}
\maketitle

\begin{abstract}
  \textbf{Introduction:} The increasing availability of large-scale textual corpora has opened new possibilities for data-driven, quantitative approaches to historical analysis using Natural Language Processing (NLP). However, diachronic corpora with historical relevance from the pre-digital era remain scarce and often incomplete.

\textbf{Methods:} We present a quantitative approach to historical analysis based on the reconstruction and exploration of a diachronic corpus of around 600,000 articles from the Italian newspaper "La Repubblica", covering all the articles published from the first of January 1985 to the 31st of December 2000 - a period of major political, social, and geopolitical change in Italy and globally.
Using NLP techniques, we analyze the text at both lexical and semantic levels; we then apply tools from complex systems and statistical physics to trace shifts in media discourse over time.

\textbf{Results:} The word usage in the corpus is bursty, with a heavy-tailed distribution of Fano factors that is robust across timescales. Changepoints in word salience and the block structure of semantic similarity allow us to detect key transition periods, such as the transition from the First Republic to the Second Republic in Italy, or major international conflicts like the Gulf War or the Kosovo War, without relying on prior labeling. Periods of war correspond to lower entropy in the semantic space, indicating a more focused discourse.

\textbf{Discussion:} The results show how combining computational linguistics with ideas from complex systems can offer new quantitative insight into historical changes, opening up new paths for studying the dynamics of media and society through large-scale textual data.
 \end{abstract}

\section{Introduction}
\label{sec:intro}
Historical research has been profoundly transformed by the advent of digital technologies. The digitization of vast archives of historical materials has enabled historians to establish the framework of "Historical Information Science" \citep{boonstraPresentFutureHistorical2004} providing a methodological lens through which to leverage the unique peculiarities of historical data. As written documentation remains the core of the work of historians, the availability of large-scale digitized textual corpora is crucial for the development of new methodologies in historical research that focus both on the transformation of documents via Optical Character Recognition (OCR) and the subsequent application of computational methods to analyze these corpora \citep{romeinStateFieldDigital2020}.

In this context, the digital age has brought significant changes, as the increasing availability of data has opened new possibilities for data-driven, quantitative approaches to the study of historical and cultural change \citep{kitchinBigDataNew2014,romeinStateFieldDigital2020}. One of the most influential examples in this regard is the field of "Culturomics" \citep{aidenUnchartedBigData2014}, which was kickstarted by the quantitative analysis of the Google Books corpus to explore  macroscopic patterns in human thought over centuries \citep{michelQuantitativeAnalysisCulture2011}. This "Distant Reading" approach \citep{morettiDistantReading2013} uses computational methods to analyze large corpora of texts, revealing patterns and trends that would be difficult to discern through traditional close reading.

From a technical standpoint, this general framework of "Content Analysis" \citep{krippendorffContentAnalysisIntroduction2018,riffeAnalyzingMediaMessages2023} has been enabled by the development of Natural Language Processing (NLP) techniques, which allows one to analyze large collections of unstructured text in a quantitative way \citep{khuranaNaturalLanguageProcessing2023}: notable NLP applications include topic modeling \citep{bleiTopicModels2009,grootendorstBERTopicNeuralTopic2022}, word embeddings\citep{mikolovEfficientEstimationWord2013,kusnerWordEmbeddingsDocument2015}, sentiment analysis \citep{wankhadeSurveySentimentAnalysis2022}, named entity recognition \citep{liSurveyDeepLearning2022}. Depending on the corpus and the research question, these techniques have been applied in linguistics \citep{goncalvesCrowdsourcingDialectCharacterization2014}, psychology \citep{bollenHistoricalLanguageRecords2021}, sociology \citep{bollenTwitterMoodPredicts2011} and more.

A particular point of view in this transition is that of complex systems: concepts from statistical physics can be borrowed to explain some peculiarities of human language and collective human dynamics by comparing them with other physical systems. The most famous example for text is Zipf's law, which demonstrates that word frequency ranking obey a power law \citep{newmanPowerLawsPareto2005}, a regularity found across biological and physical processes. Beyond frequency, human language exhibits other scaling regularities, such as "burstiness" - sudden increases in the usage of content words followed by lulls \citep{gohBurstinessMemoryComplex2008, altmannWordFrequencyBursts2009}.

From a historiographical perspective, newspapers are particularly valuable sources of information, as they provide a continuous record of events and public discourse over time \citep{leetaruCulturomics20Forecasting2011,flaounasResearchMethodsAge2013}.
While for media in the digital age there is an abundance of data available for analysis, both for newspapers and social media \citep{bakerWhoBenefitsWhen2015, partingtonModernDiachronicCorpusAssisted2010,patodkarTwitterCorpusSentiment2016,flaounasResearchMethodsAge2013,fraxanetDecadeNewsForum2026}, finding diachronic corpora from the pre-digital age is more difficult. Nevertheless, archives of historical news can be digitized and used to reconstruct the media discourse of past eras. A notable example is the analysis of the Pennsylvania Gazette, a major colonial U.S. newspaper from 1728-1800 \citep{newmanProbabilisticTopicDecomposition2006}, where topic modeling was used to uncover historical trends in eighteenth-century America; other examples include the analysis of British periodicals from 1800-1950 \citep{lansdall-welfareContentAnalysis1502017}, the exploration of Finnish newspapers from 1854-1917 \citep{marjanenTopicModellingDiscourse2021}, the exploration of Italian culture in the US through Italian-American newspapers from 1867-1920 \citep{violaMediaConstructionItalian2019} and the measurement of geographical biases in Soviet newsreels \citep{tammCityRepresentationSoviet2026}.
For Italian-language newspaper resources, relevant references from the pre-digital age include the "La Repubblica" corpus \citep{baroniIntroducingRepubblicaCorpus2004} and the corpus of the newspaper "L'Unità" \citep{basileDiachronicItalianCorpus2020}.

In this work we reconstruct the "La Repubblica" corpus from the original CD-ROM archives published in 2001, recovering around 600,000 articles covering all the issues published from 1 January 1985 to 31 December 2000, one of the most interesting periods of Republican Italian history: the final years of the \emph{First Republic} - the political system established after the World War two and dominated by the same party coalitions for decades - and the political crisis that led to the emergence of a reconfigured system commonly referred as the \emph{Second Republic} \citep{ginsborgItalyItsDiscontents2003}. We then use NLP techniques to analyze the text at both the lexical and semantic levels. The objective is to detect how historical "turning points" are reflected in the media discourse: by modeling the entire media environment as an evolving complex system, we aim to identify periods of stability and the abrupt transitions, or "regime shifts", that separate them.

The approach analyzes the corpus at multiple levels of time and content granularity: we first analyze the corpus from a statistical perspective, studying the general statistical properties of the documents; we then explore the lexical level using time series of word "salience", considering how peculiar words rise and fall in importance over time as historical events unfold, and analyze the semantic level by representing articles in a reduced semantic space using Latent Semantic Analysis (LSA) \citep{deerwesterIndexingLatentSemantic1990}, and tracking the evolution of the media discourse as a trajectory in this lower-dimensional space. By monitoring the movement of the "center of mass" of document vectors over time, we identify periods of significant change in the discourse, and the semantic dimensions most responsible for these transitions. Finally, we show how periods of war correspond to a lower entropy in the semantic space, indicating a more focused discourse.  
\section{Data}
\label{sec:data}
\subsection{The "La Repubblica" Corpus}
\label{subsec:corpus}
"La Repubblica" is one of Italy's major daily newspapers, founded in 1976 and published continuously ever since. Characterized by its reformist and progressive orientation, the newspaper has documented the country's political events, social changes, and cultural developments through political and economic analysis, investigative journalism and influential socio-cultural debate. In 2001, a set of multimedia CD-ROMs was released, containing the complete archive of \emph{La Repubblica} articles from 1 January 1985 to 31 December 2000. These discs stored the articles in a proprietary binary format, designed to be accessed through a graphical interface.

By reverse-engineering this format, we extract the full text of approximately 600,000 articles, along with metadata fields (title, publication date, section, author when available, and page number). The corpus contains approximately 300 million tokens and about 800,000 types. As the archive comes from an official source, the data quality is high and largely unaffected by OCR artefacts, which are common in digitized historical newspapers. Moreover, the corpus is mono-source (a single newspaper), which reduces stylistic variation due to editorial conventions and provides dense temporal coverage of the period.

Reconstructing the archive directly from the original CD-ROM source gives access to full article text (not available in the publicly available version of the corpus \citep{baroniIntroducingRepubblicaCorpus2004}), complete issue coverage for 1985-2000, and article-level temporal metadata suitable for fine-grained historical tracking. This structure supports the multilevel lexical-semantic analyses developed in the following sections.

While some metadata fields are missing or incomplete (e.g., author names are often absent), the publication date is always present, allowing for precise diachronic analysis. Where possible, some metadata are reconstructed from the published version of the newspaper available online. \autoref{tab:dataset_description} summarizes the dataset, showing the available columns, their completeness, and the mean token count for textual fields. The "text" field contains the full body of the article, while "headline" and "summary" provide shorter textual descriptions. The "section" field indicates the newspaper section (e.g., politics, economy, culture), which is missing in about 19\% of articles. Overall, the dataset provides a rich resource for analyzing the evolution of media discourse over a critical period in Italian history.

\begin{table}[htbp]
  \centering
  \resizebox{\textwidth}{!}{\begin{tabular}{@{}llrrrr@{}}
      \toprule
      \textbf{Category}                & \textbf{Field} & \textbf{Available} & \textbf{Missing (\%)} & \textbf{Mean Tokens} & \textbf{Mean Types} \\ \midrule
      \multirow{3}{*}{Metadata} & date         & 593593 & 0.00  & --     & --     \\
      & article page & 593593 & 0.00  & --     & --     \\
      & section      & 481748 & 18.85 & --     & --     \\ \midrule
      \multirow{5}{*}{Textual content} & headline       & 479145             & 19.28                 & 9.86                 & 9.55                \\
      & title        & 593583 & 0.00  & 6.61   & 6.50   \\
      & summary      & 209962 & 64.63 & 22.99  & 21.11  \\
      & author       & 371398 & 37.43 & 2.77   & 2.76   \\
      & text         & 593588 & 0.00  & 499.18 & 287.70 \\ \bottomrule
    \end{tabular}}
  \caption{Dataset structure and completeness. Fields are grouped by category; mean token counts and mean type counts are reported for textual fields only.}
  \label{tab:dataset_description}
\end{table}
 
\subsection{Pre-processing}
\label{subsec:pre-processing}

As a first step, the raw text of each article is pre-processed to prepare it for analysis. We apply tokenization and lowercasing, remove punctuation and stopwords\footnote{Stopwords from \url{https://github.com/stopwords-iso/stopwords-it/blob/master/stopwords-it.json}}, and perform lemmatization to reduce inflectional variability. For these steps we use the \emph{spaCy} library \citep{inesmontaniExplosionSpaCyV3722023}, which provides robust tools for Italian text processing.

\subsection{Corpus statistics}
\label{subsec:corpus_statistics}

\begin{figure}[tb]
  \centering
  \begin{subfigure}[t]{0.48\textwidth}
    \centering
    \includegraphics[width=\linewidth]{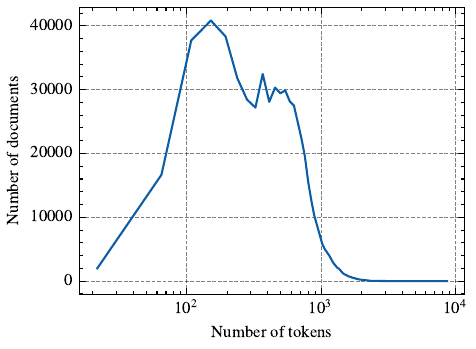}
    \caption{Distribution of the number of tokens per document.}
    \label{fig:dist_tokens}
  \end{subfigure}
  \hfill
  \begin{subfigure}[t]{0.48\textwidth}
    \centering
    \includegraphics[width=\linewidth]{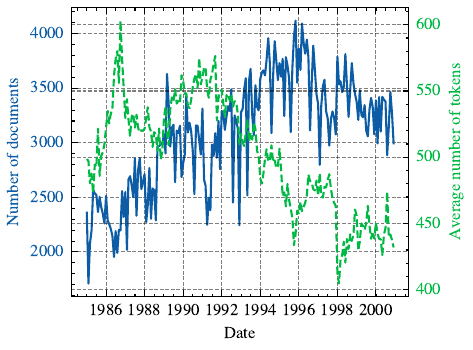}
    \caption{Number of articles and average token count per article per month during the whole period.}
    \label{fig:dist_count}
  \end{subfigure}
  \begin{subfigure}[t]{0.48\textwidth}
    \centering
    \includegraphics[width=\linewidth]{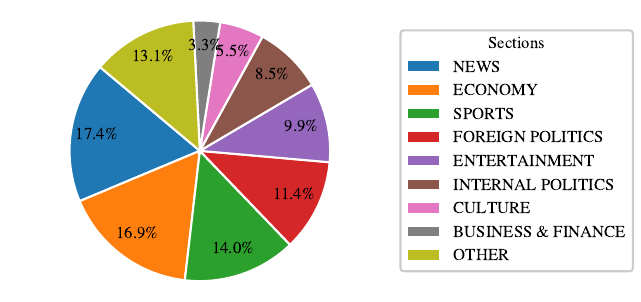}
    \caption{Composition of the corpus by section.}
    \label{fig:dist_sections}
  \end{subfigure}
  \caption{Distribution of the number of tokens per document, number of articles and average token count per article per month during the whole period, and composition of the corpus by section.}
  \label{fig:dist}
\end{figure}

To understand the general properties of the corpus we compute various statistical measures: the distribution of article lengths (in tokens), the distribution of unique words (types) per article, the main sections covered, and the temporal distribution of articles over the years. These statistics provide a baseline understanding of the corpus structure and content.  \autoref{fig:dist_tokens} shows the distributions of document length (tokens). The bimodal structure suggests heterogeneity in the corpus composition, plausibly reflecting the coexistence of short formats (e.g., brief news items) and longer forms (e.g., reports, editorials). \autoref{fig:dist_count} shows the monthly distribution of articles over the 16-year period, in terms of article count and average tokens per article. While the number of articles increases, the article length decreases. This finding is consistent with the results about "tabloidization of the news" reported in literature \citep{esserTabloidizationNewsComparative1999, mclachlanTabloidizationBritishPress2000}.
As a generalist newspaper, the composition of the corpus is diverse in terms of sections, as shown in \autoref{fig:dist_sections}.

\begin{figure}[tb]
  \centering
  \begin{subfigure}[t]{0.48\textwidth}
    \centering
    \includegraphics[width=\linewidth]{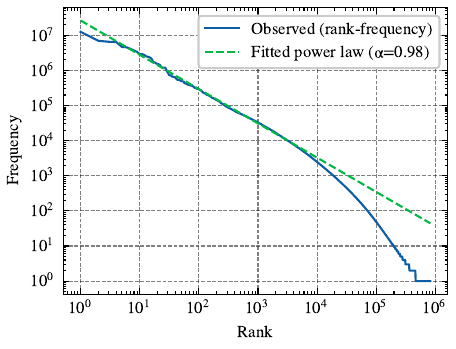}
    \caption{Rank-frequency distribution of words in the corpus (Zipf's law).}
    \label{fig:zipf}
  \end{subfigure}
  \hfill
  \begin{subfigure}[t]{0.48\textwidth}
    \centering
    \includegraphics[width=\linewidth]{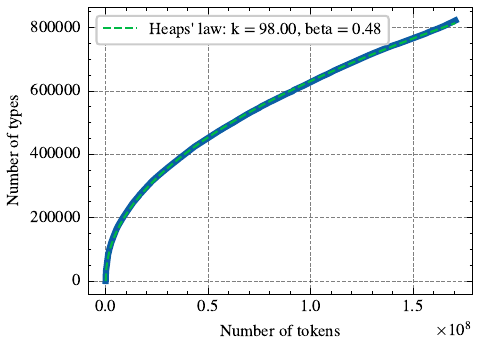}
    \caption{Growth of the vocabulary with corpus size (Heaps' law).}
    \label{fig:heaps}
  \end{subfigure}
  \caption{Empirical confirmation of Zipf's and Heaps' laws in the corpus.}
  \label{fig:zipf_heaps}
\end{figure}

The corpus adheres to the main statistical laws of natural language: Zipf's law and Heaps' law. Zipf's law \citep{newmanPowerLawsPareto2005,altmannStatisticalLawsLinguistics2016} describes the relationship between token frequency and rank in the frequency distribution: when word types are ranked in descending order of token frequency, the token frequency $f$ of a type is inversely proportional to its rank $r$, i.e., $f(r) \propto 1/r^\alpha$, usually with $\alpha\approx 1$. This heavy-tailed distribution confirms that a small number of types account for the vast majority of token usage, while most types remain rare. As shown in \autoref{fig:zipf}, in this corpus the frequency decay closely follows a Zipf's law with $\alpha \approx 1$ until a critical rank of approximately $10^4$; beyond this threshold, frequency drops off more steeply, reflecting rare types such as neologisms, technical terms, morphological variants, and typographical errors \citep{ferrericanchoTwoRegimesFrequency2001}. Relatedly, \citet{williamsTextMixingShapes2015} show that the deviation from the simple Zipf's law can also arise from the aggregation of heterogeneous texts. In a corpus comprising ~600,000 articles spanning 16 years, written by thousands of different authors and covering a broad range of topics (politics, economy, culture, sports, and more), the superposition of many different word-frequency distributions leads to a compound effect that curves the aggregate distribution down at the tail. Heaps' law \citep{altmannStatisticalLawsLinguistics2016} relates the number of distinct word types $V(n)$ to the total number of tokens $n$ in a corpus. Empirically, it follows the relation $V(n) = k n^\beta$ where $\beta \in (0.4, 0.7)$. This sublinear growth reflects the progressive slowdown in the introduction of new words as more text is observed. In our corpus, the vocabulary growth curve follows Heaps' law with $\beta \approx 0.5$, as shown in \autoref{fig:heaps}.  As a cumulative growth process, Heaps' law is not as sensitive to the heterogeneity of the corpus as Zipf's law is. Both results are consistent with similar analyses of large corpora \citep{rosillo-rodesEntropyTypetokenRatio2025}.

\section{Results and Discussion}
\label{sec:methodsresults}
\subsection{Lexical dynamics}
\label{subsec:lexical}

To capture the lexical dynamics of the corpus over time we proceed by associating each document in the corpus (i.e. the text of each article) with a representation capturing the "importance" of each word in that document. A classical approach to do so is using the Term Frequency - Inverse Document Frequency (TF-IDF) representation.
For a corpus with $N$ documents and a vocabulary of $V$ terms, a TF-IDF matrix $M \in \mathbb{R}^{N \times V}$ is constructed, where each entry $M_{dt}$ corresponds to the "salience" of term $t$ in document $d$.

Formally, for a term $t$ in document $d$:
\begin{equation}
  \begin{split}
    \mathrm{tfidf}(t, d, D) &= \mathrm{tf}(t, d) \cdot \mathrm{idf}(t, D) \\
    \mathrm{idf}(t, D) &= \log\left(\frac{1 + N}{1 + \mathrm{df}(t)}\right) + 1
  \end{split}
  \label{eq:tfidf}
\end{equation}

where $\mathrm{tf}(t,d)$ is the \emph{term frequency} (the number of times $t$ appears in $d$), and $\mathrm{df}(t)$ is the \emph{document frequency} (the number of documents containing $t$ in the full corpus $D$).

This representation is sparse, as most words do not appear in a given document, and high-dimensional, as the vocabulary size $V$ can be very large. However, applying all the pre-processing steps described in \autoref{subsec:pre-processing} helps reduce the dimensionality and sparsity of the TF-IDF matrix. To capture only the most relevant words, we further filtered the vocabulary to include only the lemmas appearing in at least 10 different articles and at most in 50\% of the articles. After the pre-processing 140 million tokens remain in the corpus, with 140,000 unique lemmas.

The salience of each word over time is then computed similarly, using the IDF as in \autoref{eq:tfidf}, but instead of calculating the TF for each document, we group all articles published in each time period $p$ into a single document $D_p$, and then compute the TF of each word $t$ in the period $p$ as $\mathrm{tf}(t, D_p)$. This way, we obtain time series of TF-IDF values for each word in the vocabulary, capturing for each how its importance evolves over time in the media discourse \citep{heAnalyzingFeatureTrajectories2007}.

\begin{figure}[tb]
  \centering
  \includegraphics[width=0.9\textwidth]{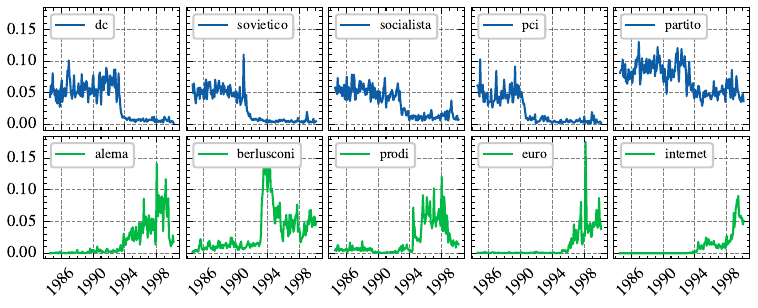}
  \caption{Monthly TF-IDF scores of falling and rising words across the time period. The top five falling words (blue) and top five rising words (green) are selected based on their linear coefficient, representing the slope of a regression line fitted to each term's monthly scores over the whole period.}
  \label{fig:top5_tfidf}
\end{figure}

To characterize the lexical dynamics we compute monthly TF-IDF time series: \autoref{fig:top5_tfidf} illustrates their heterogeneity when selecting terms with the strongest long-term positive and negative trends. These trends are quantified by the slope of a regression line fitted to the monthly scores, allowing us to identify words with the steepest rates of growth or decline over the entire period. These time series are not stationary and exhibit diverse patterns of change, both on short and long timescales, suggesting salience is driven by external events and structural transformations of the political and social landscape.

Many of the words showing a significant decreasing trend in salience are related to the political parties of the First Republic (e.g., \emph{dc}, \emph{psi}, \emph{pci}), while words related to the new political figures of the Second Republic (e.g., \emph{Berlusconi}, \emph{Prodi}) show an increasing trend. As we will below discuss in more detail, this reflects the major shift portrayed in the corpus.

A peculiar characteristic of the appearance of words is \emph{burstiness}; a measure of burstiness for counting processes, is the Fano factor $F$ is defined as
\begin{equation}
  F = \frac{\mathrm{Var}(X)}{\mathbb{E}[X]},
  \label{eq:fano}
\end{equation}
where $X$, in this case, is the TF time series of the word. A Fano factor greater than one indicates over-dispersion, characteristic of bursty behavior.

\begin{figure}[tb]
  \centering
  \begin{subfigure}[b]{0.45\textwidth}
    \centering
    \includegraphics[width=\textwidth]{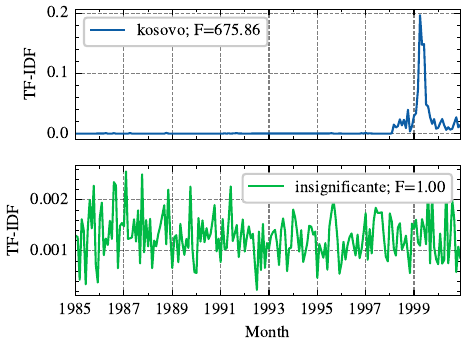}
    \caption{TF-IDF time series of words with low and high Fano factors.}
    \label{fig:low_high_fano}
  \end{subfigure}
  \begin{subfigure}[b]{0.45\textwidth}
    \centering
    \includegraphics[width=\textwidth]{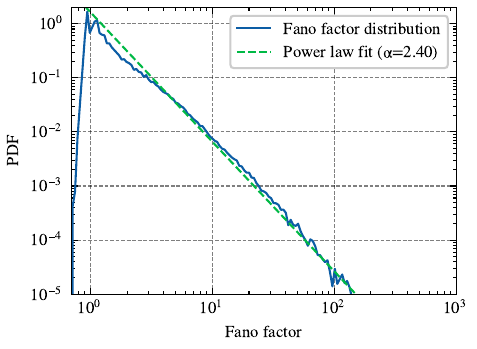}
    \caption{Fano factor distribution (log-log scale). The distribution is heavy-tailed.}
    \label{fig:fano_plot}
  \end{subfigure}\hfill

  \caption{
    Relationship between word Fano factors and temporal variability.
    (a) TF-IDF time series of two example words, illustrating a word with low Fano factor  and one with high Fano factor.
    (b) Distribution of word Fano factors (log-log scale).
  }
  \label{fig:fano_w_dist}
\end{figure}

In \autoref{fig:low_high_fano}, examples of words with high and low Fano factor are shown, while the empirical PDF of the Fano factor (\autoref{eq:fano}) is shown in \autoref{fig:fano_plot}. The distribution is heavy-tailed (exponent $\alpha=2.4$), indicating that while many terms display relatively regular dynamics, a subset shows extreme variance-to-mean ratios consistent with burst-like behavior. In substantive terms, these are typically words related to events (like \emph{kosovo} in the example), while low-Fano words are often common words (in the example \emph{insignificante}, which means "insignificant" in Italian).

\begin{figure}[tb]
  \centering
  \includegraphics[width=0.5\textwidth]{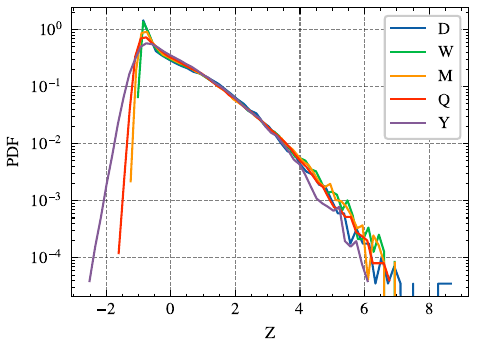}
  \caption{
  Distribution of the rescaled Fano factor across different timescales. The rescaling function is defined as $Z=(\log F - \mu_{\log F})/\sigma_{\log F}$, where $\mu_{\log F}$ and $\sigma_{\log F}$ are the mean and standard deviation of the log Fano factor at each timescale. The collapse across timescales shows that burstiness is a fundamental property of lexical dynamics in media discourse.}
  \label{fig:fano_all}
\end{figure}

Burstiness is dependent on the timescale of aggregation, reflecting that the dynamics of word usage is more or less volatile at different temporal resolutions; however, the heavy-tailed nature of the Fano factor distribution is robust across timescales, suggesting that burstiness is a fundamental property of lexical dynamics in media discourse. This is shown in \autoref{fig:fano_all}, where the Fano factor distribution is rescaled and then collapsed across different aggregation levels (day, week, month, quarter, year), indicating that the same underlying bursty dynamics govern word usage at all timescales.

\begin{figure}[tb]
  \centering
  \includegraphics[width=0.6\textwidth]{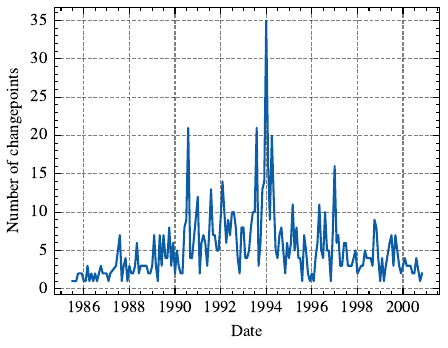}
  \caption{Number of changepoints per month, as found using the Robust PELT methodology with the L2 cost function on the monthly time series of the words in the corpus. The biggest amount of changepoints happen in the transition between the First and the Second Republic, in 1994.}
  \label{fig:changepoints_w}
\end{figure}

Finally, to identify significant shifts in word salience over time, we configure the problem as a changepoint detection task. We apply the Pruned Exact Linear Time (PELT) algorithm \citep{killickOptimalDetectionChangepoints2012} to each word's TF-IDF time series. PELT efficiently identifies points in the time series where the statistical properties change significantly, allowing us to detect moments of abrupt shifts in word usage that may correspond to important historical events or transitions in media focus. While burstiness is related to short-lived change, changepoints capture more sustained shifts in the underlying process, thus providing a complementary perspective on the dynamics of word salience.
By aggregating the detected changepoints across all words, we can identify periods of lexical shift in the discourse, and thus in the media landscape.
To evaluate the segmentation of the time series we use the L2 cost function, which measures the sum of squared deviations from the segment mean. To determine the penalty parameter $\beta$ - which controls the trade-off between fit and number of changepoints - in a data-driven way, we apply a bootstrap procedure: at each of $n$ iterations, the original time series is shuffled to destroy any temporal structure, and CROPS \citep{haynesComputationallyEfficientChangepoint2017} is run to obtain the number of changepoints $n(\beta)$ as a function of $\beta$ for the shuffled signal. After $n$ iterations, the optimal penalty $\beta^*$ is selected as the smallest value such that $\mu_{n}(\beta) + \sigma_n(\beta) < 1$, where $\mu_{n}(\beta)$ and $\sigma_{n}(\beta)$ are the mean and standard deviation of the number of changepoints detected in the shuffled data at penalty $\beta$, i.e. the smallest penalty at which the shuffled null signal is expected to yield fewer than one changepoint, ensuring that detected changepoints in the original series are unlikely to be due to noise. In \autoref{fig:changepoints_w} the algorithm is applied on the TF-IDF time series of the top 10000 words by average TF-IDF. The period with the most changepoints aligns with the transition from the First Republic to the Second Republic in the mid-1990s. This transition was triggered by a series of corruption scandals (\emph{Tangentopoli}) and the subsequent dissolution of the main political parties of the First Republic, leading to a major reconfiguration of the Italian political landscape \citep{ginsborgItalyItsDiscontents2003,koffItalyFirstSecond2000,katzElectoralReformTransformation1996,musellaPersonalLeadersParty2015}. The change of the electoral law in 1993, shifting from proportional representation to a mixed majority system, further reshaped the landscape: while the First Republic was dominated by parties, in the Second Republic individual politicians and their personalities gained more prominence \citep{katzElectoralReformTransformation1996, musellaPersonalLeadersParty2015} - a shift that the lexical trends reflect directly, with party names declining and individual politicians rising in salience. Reflecting this, the peaks in August 1993 and January 1994 are characterized by words related to the old parties (\emph{pci}, \emph{democristiano}, \emph{pli}, even the word \emph{tangentopoli}), together with words representing the general direction of the newspaper (many words related to football appear in January 1994, probably as a result of different media coverage). The transition can be understood both as a "regime change" - with the key moment in the general election of April 1994 - and as the gradual emergence of new political actors and discourses, starting from the \emph{"Mani Pulite"} investigations in the early 1990s and culminating in the 1994 elections \citep{ginsborgItalyItsDiscontents2003,koffItalyFirstSecond2000}. The peak in April 1994 is represented by members of the new government coalition (\emph{maggioranza}, \emph{maroni}, \emph{pivetti}).

Secondary peaks align with the Gulf War (\emph{persico}, \emph{arabia}) and the Germany reunification (\emph{germania}, \emph{tedesco}) both in 1990, and the economic reforms toward the Euro in 1997 (\emph{rendimento}, \emph{imposta}, \emph{lira}).

\subsection{Semantic Analysis}
\label{subsec:semantic}

To analyze the semantic content of the articles, it is necessary to capture co-occurrence structure beyond individual words. While more advanced approaches exist (e.g., contextual embeddings \citep{reimersSentenceBERTSentenceEmbeddings2019}), we use a classical approach based on Singular Value Decomposition (SVD), typically referred to as Latent Semantic Analysis (LSA) \citep{deerwesterIndexingLatentSemantic1990}. This choice provides a computationally efficient representation and supports direct inspection of the latent components. While classical LSA is typically presented on raw term-frequency counts, weighting the term-document matrix before the SVD has been standard practice since the earliest applications to information retrieval \citep{dumaisImprovingRetrievalInformation1991}. Here we use TF-IDF weights, which down-weight high-frequency terms that carry little semantic information, resulting in a TF-IDF-weighted LSA.

Starting from the TF-IDF matrix $M \in \mathbb{R}^{N \times V}$ described in \autoref{eq:tfidf}, we compute a truncated SVD,
\begin{equation}
  M \approx U_k \Sigma_k V_k^{\top},
  \label{eq:svd}
\end{equation}
where $U_k \in \mathbb{R}^{N \times k}$ and $V_k \in \mathbb{R}^{V \times k}$ have orthonormal columns, and $\Sigma_k \in \mathbb{R}^{k \times k}$ is diagonal with the $k$ largest singular values. Each document $i$ is then represented by a dense vector $\mathbf{x}_i \in \mathbb{R}^{k}$ given by the $i$-th row of $U_k\Sigma_k$. The $k$ latent dimensions in \autoref{eq:svd} can be interpreted as linear combinations of words (latent semantic components) and often capture broad semantic contrasts in the corpus.

Going from individual articles to a time-dependent description of the discourse, we aggregate document vectors by time period. For each time period $p$, let $\mathcal{I}_p$ be the index set of documents published in that time period; we define the media discourse vector or MD vector as the center of mass
\begin{equation}
  \boldsymbol{\mu}_p = \frac{1}{|\mathcal{I}_p|} \sum_{i\in\mathcal{I}_p} \mathbf{x}_i.
  \label{eq:mdvec}
\end{equation}
The sequence $\{\boldsymbol{\mu}_p\}_{p=1}^{P}$ defines a trajectory in the $k$-dimensional semantic space.

\begin{figure}[tb]
  \centering
  \includegraphics[width=0.6\textwidth]{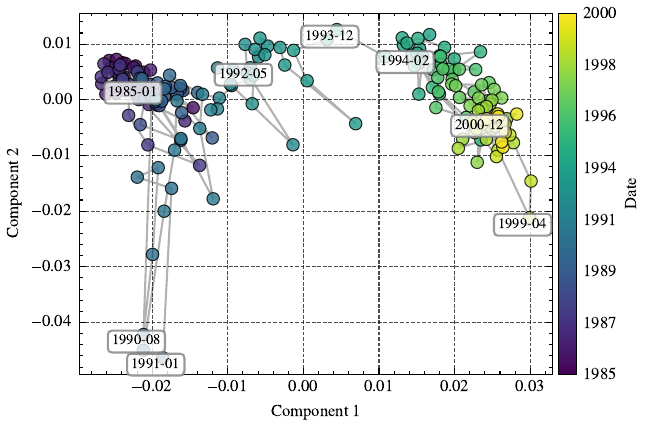}
  \caption{Trajectory of the Media Discourse projected on the two principal components. The color indicates the time.}
  \label{fig:com_trajectory}
\end{figure}

\autoref{fig:com_trajectory} shows the evolution of the monthly media discourse vector (in this case with $k=100$) projected on the first two latent components: the first component captures a political axis, related to the First and Second Republics \citep{ginsborgItalyItsDiscontents2003} - parties from the First Republic (\emph{dc}, \emph{pci}, \emph{psi}) are the words most representative for the negative values and politicians from the Second Republic (\emph{Berlusconi}, \emph{Prodi}) are the words most representative for positive values - while the second component shows a "economic stability vs. conflict" axis, with words related to economy (\emph{agreement}, \emph{stocks}) on the positive side and words related to war and conflict (\emph{war}, \emph{Iraq}) on the negative side, specifically capturing the Gulf War in the earlier period and the Kosovo War in the later period.

The resulting path is not a random walk: it contains periods of relatively smooth drift and sharper turns. This structure suggests that discourse evolution is characterized by quasi-stable regimes separated by transitions.

\begin{figure}[tb]
  \centering
  \includegraphics[width=\textwidth]{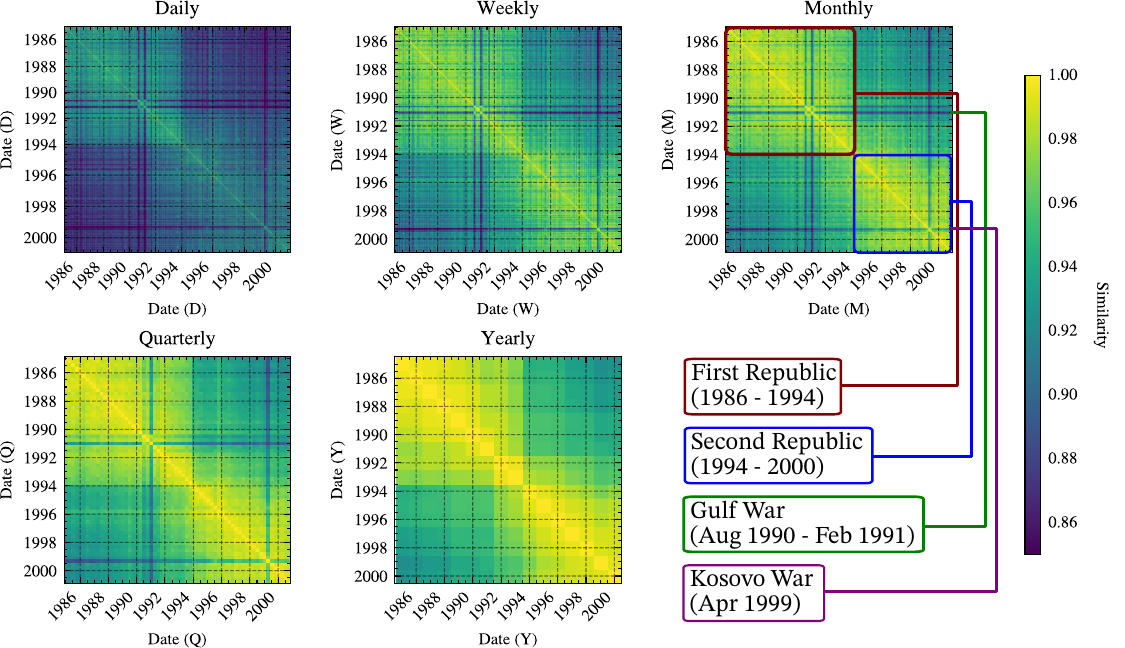}
  \caption{
    Cosine similarity matrices of the semantic center of mass at different aggregation levels, see \autoref{eq:cosine}. Darker colors indicate greater semantic change. The values are cutoff at 0.8 for better comparison across timescales.
    The block-diagonal structure becomes more apparent at coarser aggregation levels, revealing temporally coherent discourse regimes and transitions aligned with major historical events: the transition from the First to the Second Republic in 1994, the Gulf War in 1990-1991, and the Kosovo War in 1999.
  }
  \label{fig:sim_mat}
\end{figure}

The "degree of change" between time periods can be estimated via the cosine similarity between MD vectors, as it is usually done in NLP to measure the similarity between text representations:
\begin{equation}
  \mathrm{sim}(p,q)=\frac{\boldsymbol{\mu}_p\cdot\boldsymbol{\mu}_q}{\|\boldsymbol{\mu}_p\|\,\|\boldsymbol{\mu}_q\|}.
  \label{eq:cosine}
\end{equation}
Computing $\mathrm{sim}(p,q)$ for all pairs yields a similarity matrix whose block structure highlights temporally coherent regimes separated by abrupt transitions, as shown in \autoref{fig:sim_mat} over different aggregation levels.

The block-diagonal structure indicates temporally coherent segments in which discourse remains semantically similar, separated by boundaries of lower similarity. These boundaries align with historical periods, including the early-1990s transition in Italian domestic politics \citep{ginsborgItalyItsDiscontents2003,koffItalyFirstSecond2000,katzElectoralReformTransformation1996,musellaPersonalLeadersParty2015}  and major international conflicts.
The structure is robust to the choice of aggregation level; however, finer aggregation reveals more detailed transitions, while coarser aggregation highlights broader regimes. This suggests a hierarchical organization of discourse dynamics, with nested regimes and transitions at different timescales.

The phenomenon is analogous to the Epps effect in quantitative finance \citep{eppsComovementsStockPrices1979,bonannoNetworksEquitiesFinancial2004}. In financial microstructure, the empirical correlation between two assets artificially decays at high sampling frequencies due to asynchronous trading. Here, news topics do not appear synchronously, but evolve over time: for this reason, at daily aggregation, the semantic space appears noisy and fragmented, with low cosine similarity between daily MD vectors. As the aggregation window expands this noise is smoothed out, allowing the longer and stabler phases to emerge.

\subsection{Semantic diversity and entropy}
\label{subsec:entropy}

The diversity of semantic content within each time period can be quantified by measuring the dispersion of document vectors around $\boldsymbol{\mu}_p$ of \autoref{eq:mdvec}. Let $\Sigma_p$ denote the empirical covariance matrix of $\{\mathbf{x}_i\}_{i\in\mathcal{I}_p}$ and $\{\lambda_{p,r}\}_{r=1}^{k}$ its eigenvalues.  The \emph{spectral entropy} is defined on the normalized eigenvalue spectrum $\pi_{p,r}=\lambda_{p,r}/\sum_{s=1}^{k}\lambda_{p,s}$ as
\begin{equation}
  H(p)=-\sum_{r=1}^{k}\pi_{p,r}\log \pi_{p,r}.
  \label{eq:entropy}
\end{equation}
Lower entropy values indicate a contraction of semantic diversity, consistent with a discourse that focuses on a narrower set of themes. A contraction in the eigenvalue spectrum of the covariance matrix indicates that discourse variability collapses onto fewer semantic directions: articles become more similar in content and framing, even when they are not identical. Importantly, this measure is insensitive to the specific vocabulary used; it captures discourse focusing in a representation defined by co-occurrence structure.

\begin{figure}[tb]
  \centering
\begin{subfigure}[t]{0.48\textwidth}
    \centering
    \includegraphics[width=\textwidth]{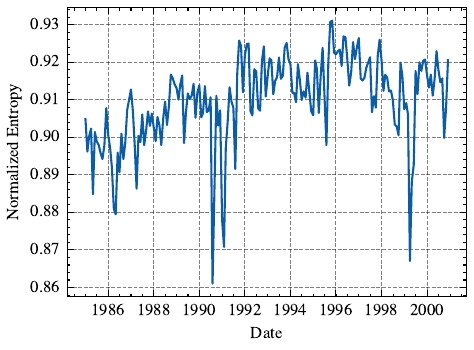}
    \caption{Monthly (M) entropy estimation.}
    \label{fig:entropy_m}
  \end{subfigure}
  \hfill \begin{subfigure}[t]{0.48\textwidth}
    \centering
    \includegraphics[width=\textwidth]{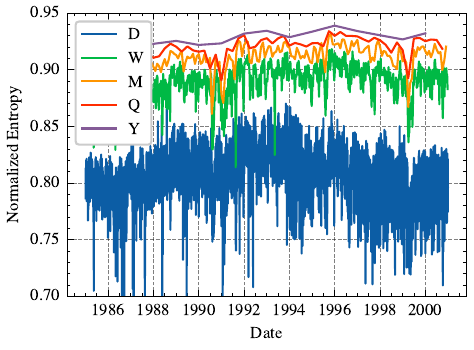}
    \caption{Comparison across all scales (D, W, M, Q, Y).}
    \label{fig:entropy_all}
  \end{subfigure}

  \caption{Comparison of spectral entropy estimations, see \autoref{eq:entropy}. The left panel shows high-frequency monthly fluctuations, while the right panel illustrates how entropy stabilizes as the aggregation window increases.}
  \label{fig:entropy_comparison_side_by_side}
\end{figure}

\autoref{fig:entropy_m} shows a contraction of semantic spread in correspondence with crisis periods: during major conflicts, the month-level distribution in semantic space becomes more anisotropic, with variance concentrating in fewer directions, consistent with agenda compression \citep{boydstunMakingNewsPolitics2013}.

\autoref{fig:entropy_all} demonstrates how this semantic diversity scales with the aggregation window. At high frequencies (Daily), entropy remains low and highly volatile, reflecting the sparse, event-driven nature of daily publications. As the temporal window increases (Weekly to Yearly), the entropy rises and stabilizes, illustrating a return to a broad, isotropic topic distribution.

One of the main topics during low-entropy periods is war and conflict, which tends to dominate media attention and reduce the diversity of covered topics. In particular, the entropy minima align with Operation Desert Shield (Gulf War, August 1990), Operation Desert Storm (Gulf War, February 1991), and the Kosovo War (April 1999). During these extreme crisis periods, the agenda compression is severe enough to cause sharp drops in entropy across almost all aggregation scales, temporarily overpowering the natural temporal smoothing of the semantic space.

These contractions are more evident at an intermediate aggregation level. At daily resolution, each window contains the articles of a single issue, each very heterogeneous in topics: there are standing sections (politics, economy, culture, sports) and there are some leading stories. The leading story changes from day to day: this means that the daily entropy fluctuates but, in ordinary times, no contraction persists. At yearly resolution, a war lasting 2-3 months leads the agenda for only a fraction of the window, and its contraction is averaged out. The monthly aggregation is short enough that an episode holding the lead for weeks appears as a coherent contraction, and long enough to smooth out the daily rotation of leading stories.
 
\section{Conclusions}
\label{sec:conclusions}
The results support a two-level description of diachronic media discourse. At the lexical level, word salience exhibits strong heterogeneity: some terms change gradually, whereas others display burst-like dynamics with sudden spikes and rapid decay. The heavy-tailed distribution of Fano factors indicates that such intermittency is a systematic feature of specific vocabulary, consistent with the idea that exogenous events intermittently restructure attention. This burstiness is observed across different timescales, suggesting that the dynamics of word usage is governed by similar underlying processes at multiple temporal resolutions. The invariance of the distribution's shape under rescaling suggests that the process belongs to a scaling class, a hypothesis that could be explored in future research.

At the semantic level, the center-of-mass trajectory and the cosine similarity matrix suggest that these local lexical fluctuations aggregate into mesoscopic regimes. Word-level bursts
are not noise: when they align across many terms, they produce an observable reorganization of the discourse as a whole, which is representative of the events of the period in consideration. The block-diagonal similarity structure provides a compact representation of this phenomenon, turning a sequence of heterogeneous texts into a map of quasi-stable phases and transitions.

The mid-1990s emerge as the most significant period of change in the corpus. As a newspaper, a large part of the coverage is naturally focused on political events, and the transition from the First Republic to the Second Republic in Italy is probably the most important turning point in Republican Italy political history \citep{ginsborgItalyItsDiscontents2003}. This is reflected in the semantic trajectory, which starts drifting away from the First Republic cluster in the early 1990s and stabilizes in a new region of the semantic space in 1994, while the similarity matrix shows a clear boundary in April 1994.

Beyond the domestic political transition, the entropy analysis provides quantitative support, in this corpus, for a recurring qualitative claim in studies of media and crises: exceptional events compress the agenda, producing a narrower set of salient themes \citep{iyengarNewsCoverageGulf1993,boydstunMakingNewsPolitics2013}. The Gulf War and the Kosovo War are the most salient examples of this phenomenon in the corpus, with sharp drops in semantic diversity across all aggregation scales. The same periods appear in the similarity matrix as points that are very different from the months before and after.

The choice of LSA reflects a deliberate methodological trade-off.
While LSA components remain linear combinations of words and
cannot capture context-dependent meanings as well as more advanced
embedding methods, they offer easy interpretability by lexical
inspection - for instance, examining the terms with largest
loadings in the components responsible for major turns in the
trajectory. This may be particularly relevant for historical
corpora where the priority is to reconstruct and interpret
macro-dynamics rather than optimize predictive performance. In the
present setting, the central object of interest is the temporal structure of change (regimes and transitions). The observed block-diagonal similarity structure and the entropy contractions are robust, high-level patterns that do not require a highly nonlinear representation to emerge. Crucially, the approach is entirely unsupervised, requiring no prior labeling of historical periods, which makes it directly applicable to corpora where such labels are absent or contested. A possible extension of our work could be to perform a discourse analysis, with the aid of context-dependent embeddings, to track how the newspaper positioned itself in the debates of the period investigated here.

Some limitations qualify these conclusions.
First, the corpus is mono-source, reflecting the editorial priorities and stylistic conventions of a single newspaper, though "La Repubblica" wide national circulation makes it a reasonable proxy for dominant public discourse. Moreover, the corpus is bounded to 1985-2000. Extending the analysis to later years or comparing multiple
newspapers would allow testing of the persistence and generality of the observed patterns and to capture subsequent political and social transformations. The present approach is also intentionally unsupervised and global; a topic-based analysis, using for example Latent Dirichlet Allocation or dynamical topic models, could complement it by zooming in on specific themes, making it interesting for further research. Within these limits, however, the study illustrates a transferable workflow compatible with historical inquiry: reconstruct a diachronic corpus; extract lexical time series to identify shifts in salient vocabulary; apply semantic analysis to map discourse change systematically; and identify historical turning points as regime shifts in the semantic space.
The detected transitions can then be interpreted in light of
historical knowledge, and the semantic dimensions responsible for
these transitions can be analyzed to understand the underlying
themes and narratives. The approach is not meant to replace
traditional historical analysis, but rather to complement it by  providing a data-driven perspective on the dynamics of media and society.
 
\section*{Conflict of Interest Statement}
The author(s) declared that this work was conducted in the absence of any commercial or financial relationships that could be construed as a potential conflict of interest.

\section*{Generative AI Statement}
The author(s) declared that generative AI was not used in the creation of this manuscript.

\section*{Author Contributions}
DZ: Validation, Formal Analysis, Conceptualization, Writing -- review and editing, Methodology, Data curation, Software, Writing -- original draft, Visualization, Investigation. SM: Conceptualization, Investigation, Funding acquisition, Writing -- review and editing, Validation, Writing -- original draft, Project administration, Supervision, Formal Analysis, Methodology. DS: Writing -- review and editing, Validation, Writing -- original draft, Formal Analysis, Project administration, Funding acquisition, Methodology, Investigation, Conceptualization, Supervision.

\section*{Funding}
DZ acknowledges financial support in this work by the European Union - Next Generation EU through the project GRINS "Growing Resilient, INclusive and Sustainable" funded within the framework Piano Nazionale di Ripresa e Resilienza (PNRR), mission 4 "Istruzione e Ricerca" Component 2 "Dalla Ricerca all’Impresa" Investment 1.3 (PE0000018).
DS acknowledges partial support from Grant No. PID2024-157493NB-C21 funded by MI-CIU/AEI/10.13039/501100011033 and “ERDF/EU”, and from Grant No. CEX2021-001164-M funded by the Maria de Maeztu Program for Units of Excellence in R\&D, and from Grant No. PRD2025\_92 funded by CAIB.

\section*{Acknowledgments}
D.Z. conducted part of this work during a visiting period at IFISC (UIB-CSIC), whose hospitality is gratefully acknowledged. We thank Pablo Rosillo-Rodes for helpful discussions.

\section*{Data Availability Statement}
The datasets presented in this article are not readily available because they were reconstructed from proprietary archives and contain full article texts and metadata subject to publisher copyright and licensing restrictions, so they cannot be redistributed. A derived dataset has been uploaded to a publicly accessible repository: \href{https://figshare.com/articles/dataset/Daily_aggregated_TFIDF_of_top_10k_words_in_La_Repubblica_1985-2000_news_corpus/32857679?file=66177827}{\nolinkurl{figshare.com/articles/dataset/32857679}}. Requests to access the datasets should be directed to Dario Zarcone, dario.zarcone01@unipa.it.
 
\bibliographystyle{apalike-doi}

\end{document}